\documentclass[twocolumn,
superscriptaddress,
prapplied,
showpacs,
preprintnumbers,
amsmath,amssymb]{revtex4}

\usepackage{graphicx,epsf,epsfig,ulem}
\usepackage{epsfig}
\usepackage{epstopdf} 
\usepackage{bm}
\usepackage{subfigure}
\usepackage{color,soul}
\usepackage{xcolor}
\usepackage{framed}
\usepackage{amsmath}
\colorlet{shadecolor}{yellow}

\begin{document}

\title{All-electrical control of donor-bound electron spin qubits in silicon}
\textbf{}

\author{Yu Wang*}
\affiliation{Network for Computational Nanotechnology, Purdue University, West Lafayette, IN 47907, USA}

\author{Chin-Yi Chen}
\affiliation{Network for Computational Nanotechnology, Purdue University, West Lafayette, IN 47907, USA}

\author{Gerhard Klimeck}
\affiliation{Network for Computational Nanotechnology, Purdue University, West Lafayette, IN 47907, USA}

\author{Michelle Y. Simmons}
\affiliation{Centre for Quantum Computation and Communication Technology, School of Physics, University of New South Wales, Sydney, NSW 2052, Australia}

\author{Rajib Rahman*}
\affiliation{Network for Computational Nanotechnology, Purdue University, West Lafayette, IN 47907, USA}
\date{\today}

\begin{abstract}
We propose a method to electrically control electron spins in donor-based qubits in silicon. By taking advantage of the hyperfine coupling difference between a single-donor and a two-donor quantum dot, spin rotation can be driven by inducing an electric dipole between them and applying an alternating electric field generated by in-plane gates. These qubits can be coupled with exchange interaction controlled by top detuning gates. The qubit device can be fabricated deep in the silicon lattice with atomic precision by scanning tunneling probe technique. We have combined a large-scale full band atomistic tight-binding modeling approach with a time-dependent effective Hamiltonian description, providing a design with quantitative guidelines. 

\end{abstract}


\maketitle 

\section{Introduction}

Donor bound electrons in silicon have been demonstrated to have long spin coherence times \cite{Tyryshkin, Buch, Hsueh}, making them promising candidates for solid-state qubits. Few-donor quantum dots \cite{Buch, Weber_DQD} have been patterned by scanning tunneling microscopy (STM) based lithography in silicon with atomic precision \cite{Schofield}, an excellent fabrication technique for building a scalable quantum computer. Due to different quantum confinement and hyperfine interaction compared to single donors, few-donor quantum dots provide more flexibility in addressing \cite{Buch, YW_hyperfine} and engineering the exchange coupling \cite{YW_exchange}, which are favorable attributes for multi-qubit operations. 

Controlling individual electron spins is of great importance for donor-based quantum computation. Manipulation of electron spins with integrated microwave antenna has been demonstrated in both donor and gate-defined quantum dot qubits with long coherence and high gate fidelity \cite{Muhonen, Menno1}. However, it is challenging to use an ac magnetic field to realize local spin control for many qubits, a crucial requirement for multi-qubit operations in a scalable quantum computer architecture. We also know that a microwave antenna can introduce deleterious noise to the coherence of a qubit \cite{Muhonen}. An alternative way to spin control is to utilize an oscillating electric field, which has been demonstrated in quantum dot systems \cite{Kato, Laird, Vandersypen}. Here, the qubit is modulated periodically by the difference in Zeeman energy caused by either non-uniform electron g-factor, external magnetic field or hyperfine couplings \cite{Pingenot, Tokura, Petta}. To date, all-electrical control of spins without a microwave magnetic field in donor systems in silicon has not been demonstrated. 
\begin{figure}[htbp]
\center\epsfxsize=3.4in\epsfbox{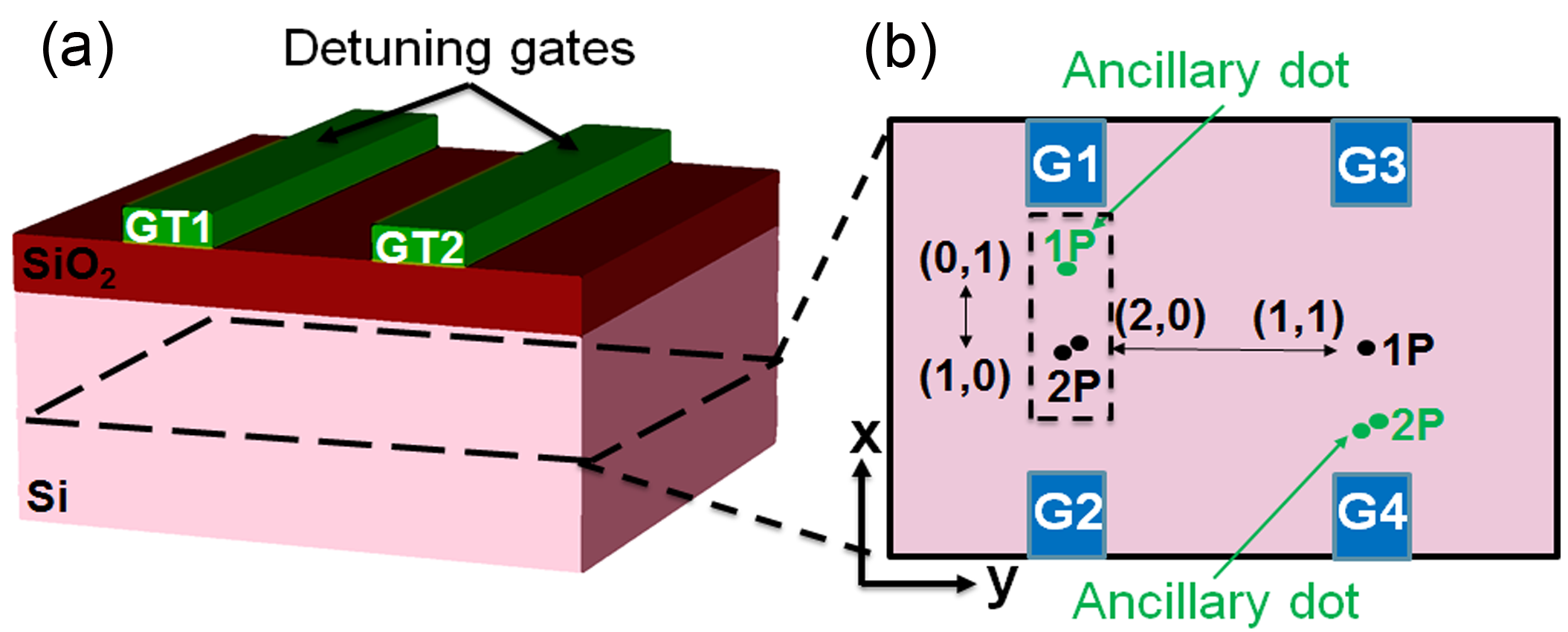}
\caption{ \textbf{Electrical control of donor-based spin qubits.} (a) The pink box denotes the silicon substrate. The two gates shown in green (GT1 and GT2) are the surface detuning gates for controlling two-qubit exchange interaction. The dashed contour indicates the plane that contains the central device components (as shown in (b)). (b) The blue rectangles represent the in-plane gates (G1-G4) to induce the (1,0) and (0,1) \cite{charge_config_note} transitions for single-qubit operations. The (2,0) and (1,1) transition is controlled by the surface gates (GT1 and GT2) in (a) for two-qubit operations.}
\vspace{0cm}
\label{fig_devices}
\end{figure}

In this work, we propose all-electrical control of donor-based spin qubits taking advantage of the hyperfine coupling difference between single donor and few-donor quantum dots by introducing ancillary dots (The device schematic is shown in Fig. 1(a) and (b) with ``P'' denoting phosphorus donors). The electron and the 3 nuclear spins (of 2P + 1P atoms) in the dashed box in Fig. 1(b) define the single-qubit operation space, and the information is encoded in the electron spin. Here the ancillary dot (1P in green) creates a difference in the local hyperfine coupling between the 1P and 2P dots due to the different number of nuclei and asymmetric quantum confinement in the dashed box \cite{YW_hyperfine}. The (1,0)$\leftrightarrow$(0,1) \cite{charge_config_note} charge transition between them can be controlled by the pair of in-plane gates G1 and G2. An electric dipole can thereby be induced by biasing the system near the (1,0)--(0,1) charge degeneracy point, where the hyperfine couplings can be modulated with an ac electric field on top of the dc electric field from the in-plane gates G1 and G2 to drive the electron spin transition. 

In this proposed approach, the donor dots can be placed with atomic precision far from interfaces or surfaces using an STM based lithography technique, making them less prone to noise sources close to the interface \cite{Shamim, Shamim2, Schenkel, Paik}. Using Coulomb confined states leads to higher valley and orbital states that are not accessible as they are typically at least 10 meV above the ground state. This results in well-isolated states of operation within which the qubit coherence can be boosted. Moreover, this scheme could be realized with existing circuitry in STM-patterned devices \cite{Weber_DQD, Tom_readout} without introducing extra control components. The donor dots can be coupled via the exchange interaction (e.g. the (1,1)$\leftrightarrow$(2,0) transition in Fig. 1(b)) controlled by the surface detuning gates GT1 and GT2 (Fig. 1(a)), retaining the highly tunable exchange coupling \cite{YW_exchange} and allowing fully electrical two-qubit operations. This design can be potentially extended to a scalable quantum computer architecture by repeating the fundamental structure (Fig.  1(a) and (b)) according to the requirements of the large-scale architecture \cite{Fowler, Hill}.

\section{Methods}
In the following, we describe the spin and charge evolution in the 2P-1P system by an effective Hamiltonian with quantitative details. The effective Hamiltonian can be expressed as:
\begin{equation}
  H = H_e + H_T + H_{Z_e} + H_{Z_n} + H_{HF} , \label{eq1}
\end{equation}
with the basis $\vert s,i_1i_2i_3,d\rangle$ including spin and charge information, where we denote $s=\uparrow,\downarrow$ as electron spin, $i_j=\Uparrow,\Downarrow$ (j = 1,2,3. See Fig. 2(a)) as the three nuclear spins, and $d=2P,1P$ as the specific dot site (e.g. $\vert \downarrow,\Uparrow\Uparrow\Uparrow,2P\rangle$ is the lowest energy configuration). $H_e$ reflects both the on-site energy detuning and the applied ac electric field, which is expressed as: 

\begin{equation}
H_e = \sum_{d=1P,2P}(\epsilon_d +e \vec{E_{ac}}\cdot\vec{R})\vert d\rangle\langle d\vert, \label{eq1-1} 
\end{equation}
where $\epsilon_d$ is the on-site detuning energy, and $d$ is the dot index (1P or 2P). $e$ is the elementary charge, $\vec{E_{ac}}$ is the ac electric field, and $\vec{R}$ is the separation between the 2P and the 1P dots. The second term represents the tunnel coupling between the (1,0) and (0,1) charge states of the 2P and the 1P dots:
\begin{equation}
H_T = \sum_{d\neq d'} t_c\vert d\rangle\langle d'\vert, \label{eq1-2}
\end{equation}
where $t_c$ denotes the tunnel coupling between the two donor dots. $d$ and $d'$ denote different dot indices. The third and the fourth term denote the Zeeman energy of the electron spin and the nuclear spins:
\begin{equation}
H_{Z_e} = g_e\mu_B\vec{B}\cdot\vec{S}, \label{eq1-3}
\end{equation}
\begin{equation}
H_{Z_n} = \sum_jg_n\mu_B\vec{B}\cdot\vec{I_j}, \label{eq1-4}
\end{equation}
where $g_e$ and $g_n$ are the electron and nuclear g-factors respectively. $\mu_B$ and $\mu_n$ are the Bohr and the nuclear magnetons respectively. $\vec{S}$ denotes the electron spin operator, and $\vec{I_j}$ denotes the spin operator of the $j$th nucleus.The fifth term gives the hyperfine coupling between the qubit electron and the donor nuclei:
\begin{equation}
H_{HF} = \sum_j A_j\vec{I_j}\cdot\vec{S}, \label{eq1-5}
\end{equation}  
where $A_j$ is the Fermi-contact hyperfine coupling between the qubit electron and the $j$th nucleus.

To provide quantitative guidelines for exploiting this proposal, it is important to obtain the parameters in $H$, the tunnel coupling $t_c$, Fermi contact hyperfine couplings $A_j$, with sufficient accuracy. Here we model the system using an atomistic tight-binding approach to obtain the Stark-shifted electron wavefunctions, from which tunnel couplings $t_c$, hyperfine couplings $A_j$ and their electric field dependency can be extracted \cite{YW_hyperfine, Rahman_prl, Park}. In the tight-binding approach, the atoms are represented by $sp^3d^5s^*$ atomic orbitals with spin-orbit coupling and nearest-neighbor interactions. Each donor is represented by a Coulomb potential screened by the dielectric constant of silicon and an on-site constant potential for the Coulomb singularity, calibrated with the P donor energy spectrum \cite{Ahmed}. The donor ground-state wavefunction obtained from this approach agrees well with the recent STM imaging experiments \cite{Salfi}. This physics-based approach automatically includes silicon conduction band valley degrees of freedom, and captures valley-orbit interaction \cite{Rahman_gfactor} and Stark effect in donor orbitals \cite{Rahman_prl}. The experimentally spin relaxation times of a single P donor and few-donor dots can also be reproduced with our approach \cite{Hsueh}. This provides us confidence in accurately extracting the parameters $t_c$ and $A_j$ under realistic electric fields. 

\section{Results and discussions}
\begin{figure}[htbp]
\center\epsfxsize=3.4in\epsfbox{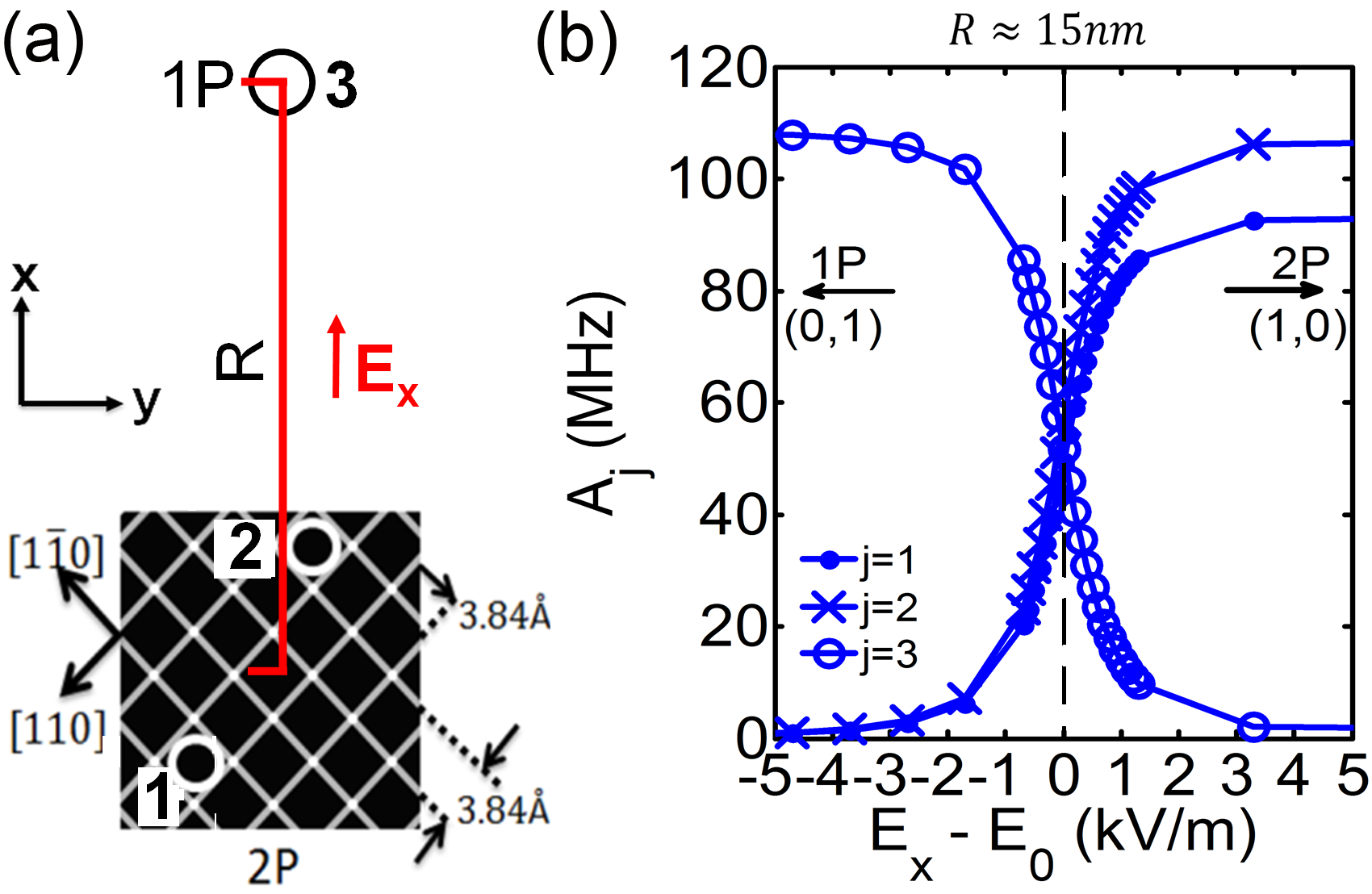}
\caption{ \textbf{Electric field dependence of the hyperfine couplings.} (a) The black squares represent the silicon atoms on a (001) atomic plane and the open circles represent the substituting P atoms. The ancillary 1P dot is placed away from the center of the 2P dot along the equivalent [010] crystallographic direction (the x direction) by $R$. The nuclei associated with the three donors in the system (in the dashed box in Fig. 1(b)) are labeled as 1, 2 and 3. (b) The hyperfine couplings associated with each nucleus as a function of electric field $E_x$ for $R\approx$15 nm. $E_0$ = 4963.3 kV/m.}
\vspace{0cm}
\label{fig_hyperfine}
\end{figure}

Fig. 2 shows the modulation of the hyperfine couplings ($A_j$) between the qubit electron spin and the nuclear spins of the P donors (2P + 1P) with electric field, with j=1,2 labeling the nuclei of the 2P dots respectively and j=3 labeling the nucleus of the 1P dot. The detail of the system in the dashed box in Fig. 1(b) is depicted as Fig. 2(a), where an atom configuration of the 2P dot is shown at the bottom. An in-plane external electric field ($E_x$) between G1 and G2 applied along the direction between the 2P dot and its ancillary 1P dot can detune the system and redistribute the electron wavefunction between them, thereby controlling the hyperfine couplings of the 2P and the 1P dots, as shown in Fig. 2(b). As can be seen, $A_3\approx 0$ at $E_x-E_0 = 5kV/m$, and $A_1, A_2\approx 0$ at $E_x-E_0 = -5kV/m$, indicating that the charge states (0,1) and (1,0) can be accessed with a small electric field range ($\sim$ 10 kV/m) with an inter-dot separation R$\approx$15 nm. At $E_x \approx E_0$, the (1,0) and (0,1) charge states are degenerate, forming hybridized bonding and anti-bonding states. As observed, the hyperfine couplings ($A_j$) have a nearly linear dependence on the electric field near the (1,0)--(0,1) charge degeneracy point ($-1 kV/m <E_x - E_0 < 1 kV/m$), indicating that the hyperfine coupling difference between the 2P and the 1P dot has a linear response to the external electric field. Near the charge degeneracy point ($E_x \approx E_0$), an electric dipole transition can be driven by an ac electric field $\varepsilon_0 sin(\omega t)$ applied from G1 and G2, thereby causing the modulations of $A_j$ with time. When the ac electric field is in resonance with the qubit energy splitting (solved from the static part of $H$ in eq. (\ref{eq1})), the electron spin transition can be induced by the overall local hyperfine coupling difference \cite{Laird, Tosi} between the 2P and the 1P dot. The emulation of this process in a 2P-1P system is now described by solving the time evolution based on the Hamiltonian in eq. (\ref{eq1}).
\begin{figure}[htbp]
\center\epsfxsize=3.4in\epsfbox{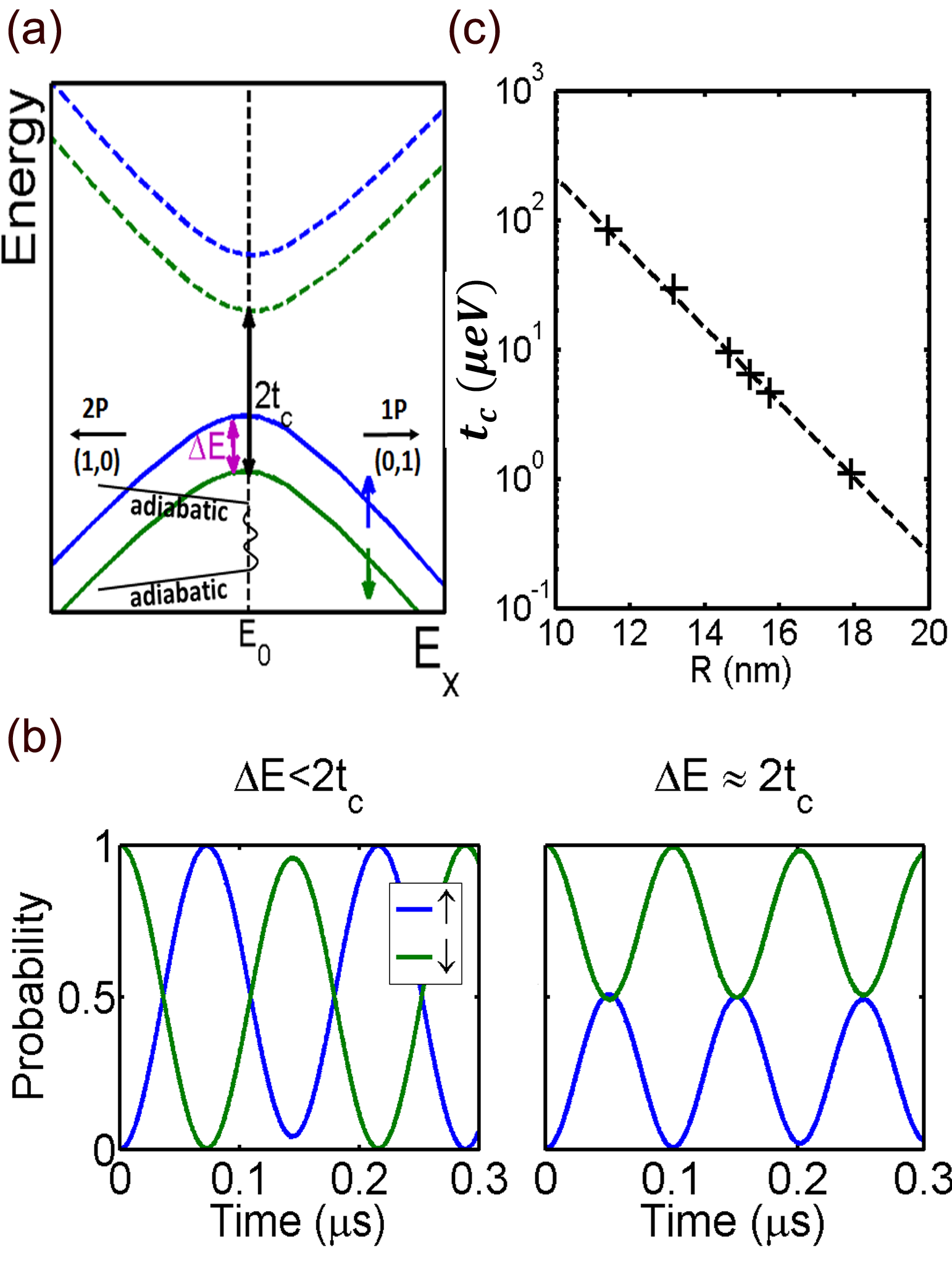}
\caption{ 
\textbf{Impact of inter-dot separation on tunnel coupling and Rabi oscillations.}
(a) The simplified energy diagram of the 2P-1P system under a static magnetic field near the (1,0)--(0,1) charge degeneracy point($E_0$). The solid green (electron spin down) and blue (electron spin up) curves represent the two bonding states that define the qubit, separated by $\Delta$E in energy. The dashed curves are the anti-bonding states, $2t_c$ above the bonding states respectively. The explicit nuclear spins are not shown here for simplicity. (b) Electron spin Rabi oscillations under $\Delta$E$<2t_c$ and $\Delta$E$\approx$2$t_c$.
(c) The tunnel coupling ($t_c$) as a function of inter-dot separation ($R$). The markers $``+"$ indicate data extracted from the atomistic tight-binding simulations, and the dashed curve is fitted to the data with the regression function $t_c = t_0e^{-bR}$, where $t_0$ = 0.1742 eV and b = 0.67 $nm^{-1}$. }
\vspace{0cm}
\label{fig_rabi}
\end{figure}

We assume the system is initially in its lowest energy configuration and the electron is located at the 2P dot (at (1,0)), i.e. $\vert\downarrow,\Uparrow\Uparrow\Uparrow,2P\rangle$, and assume the 2P dot configuration as shown in Fig. 2(a) with the 2P in one dot $\leq$1.5 nm apart. A static magnetic field $B_0$ is applied along the separation direction $x$. One of the possible ways to prepare the qubit in this initial state is to utilize the dynamic nuclear polarization technique {\cite{Petta_dynamic}} by repeatedly and selectively loading up-spin electrons to the donor dots {\cite{Morello_readout}} and emptying the dots to dynamically drive the nuclear spins to up orientations, then depleting the dots and loading a down-spin electron onto the dots in the end.  

To achieve universal quantum gates, two-axis control of single qubits, i.e. Z-gate and X-gate, is needed. A Z-gate can be simply realized by applying the external static B-field. We will focus on the X-gate in the following. As shown in Fig. 3(a), at the beginning of the control manipulation, an adiabatic in-plane gate bias between G1 and G2 is applied to ramp the static dc electric field up to and to keep it at $\sim$$E_0$, which is followed by a continuous ac electric field between the same gates. The X-gate rotation (the manipulation of $\downarrow\rightarrow\uparrow$ or $\uparrow\rightarrow\downarrow$) is then driven by the ac electric field if $h\gamma = \Delta E$, where $\Delta E$ is the energy difference between the two qubit states and $\gamma$ is the frequency of the ac electric field. To be explicit, the qubit ground state is $\frac{1}{\sqrt{2}}(\vert\downarrow,\Uparrow\Uparrow\Uparrow,2P\rangle - \vert\downarrow,\Uparrow\Uparrow\Uparrow,1P\rangle)$, and the qubit excited state is a dressed state involving both electron and nuclear spins, which can be expressed as $\alpha(\vert\uparrow,\Downarrow\Uparrow\Uparrow,2P\rangle - \vert\uparrow,\Downarrow\Uparrow\Uparrow,1P\rangle) +\beta(\vert\uparrow,\Uparrow\Downarrow\Uparrow,2P\rangle$
$- \vert\uparrow,\Uparrow\Downarrow\Uparrow,1P\rangle) + \zeta(\vert\uparrow,\Uparrow\Uparrow\Downarrow,2P\rangle - \vert\uparrow,\Uparrow\Uparrow\Downarrow,1P\rangle)$. A second adiabatic gate bias is then applied after the X-gate control is finished to bring the electron back to the 2P dot for spin storage. Here, we choose 2P over 1P because the electron spin relaxation time is longer in a 2P donor cluster than a single donor dot \cite{Hsueh}.

To achieve high-fidelity X-gate operation and make the system less prone to charge noise and relaxation, we need to make appropriate choices of the external B-field and the tunnel coupling. On the one hand, with regard to the external B-field, the qubit energy splitting $\Delta$E can be expressed as $E_{Z_e} + \delta$, where $E_{Z_e}$ is the electron Zeeman splitting, and $\delta$ includes the effects of nuclear spin Zeeman energies and hyperfine couplings, which contributes to an effective magnetic field in the order of mT. To form well-defined qubit states and suppress nuclear spin flip-flop, we need $E_{Z_e} >> \delta$ to preserve the qubit state when no ac field is applied. As a result, the external B-field is required to be in the order of 0.1 T. 

On the other hand, regarding the tunnel coupling, we need $\Delta$E significantly smaller than $2t_c$ in order to make the higher anti-bonding states well separated from the lower qubit states, preventing state hybridization or excitation due to environmental noise. As an example, Fig. 3(b) shows the Rabi oscillations of the qubit electron spin under the driving ac electric-field under $\Delta E<2t_c$ ($B_0$ = 0.5 T) and $\Delta E\approx 2t_c$ ($B_0$ = 1.45 T) for $R\approx$ 11.4 nm, where the magnitude of the driving ac electric field is 15 kV/m, and its frequency is $\gamma\approx$ 14 GHz which satisfies $\Delta E = h\gamma$. The ac electric field is assumed to be a single-frequency sinusoid. As shown, a full X-gate spin rotation can be achieved for $\Delta E<2t_c$, while the X-gate fidelity (defined as max($\sum\vert\langle\Psi\vert\uparrow,i_1i_2i_3,d\rangle\vert^2$)) is diminished when $\Delta E\approx 2t_c$ due to the qubit spin-up state (solid blue curve in Fig. 3(a)) is hybridized with the upper anti-bonding spin-down state (dashed green curve). As a result, $t_c$ needs to be engineered large enough. Fig. 3(c) shows $t_c$ as a function of the inter-dot separation $R$. As can be seen, $t_c$ decreases exponentially as a function of $R$, because the wavefunction overlap of the 2P and the 1P dots decreases exponentially as $R$ increases. To achieve $\Delta E<2t_c$, if we choose B = 0.1 T, $R$ needs to be larger than 15.6 nm approximately according to Fig. 3(c). If B = 0.5 T is chosen, $R$ needs to be at least 13 nm.

In the following, we investigate the effect of $R$ (or $t_c$) on the qubit coherence time. Both magnetic and charge noise can lead to qubit decoherence \cite{Kuhlmann}. In the proposed design, magnetic noise can be suppressed to a large extent if the substrate is made of enriched Si-28. In addition, the microwave antenna that introduces magnetic noise \cite{Muhonen} in the traditional magnetic qubit manipulation is excluded here. As a consequence, we mainly consider the effect of the charge noise from the charge fluctuations in the nearby gates on qubit coherence. 
We investigate the decoherence time $T_{2}^{*}$ possibly due to different types of charge noise from a single nearby in-plane gate, e.g. G1 in Fig. 1(b). $T_{2}^{*}$ can be obtained using \cite{Chirolli}:
\begin{equation}
\begin{split}
\frac{1}{T_{2}^{*}}  = \frac{e^2}{\hbar^2}\vert\sum_{r_i=x,y,z}\langle\Psi_{\uparrow}\vert r_i\vert \Psi_{\uparrow}\rangle-\langle\Psi_{\downarrow}\vert r_i\vert \Psi_{\downarrow}\rangle\vert^2\\\cdot\frac{S_E(\omega)}{\omega}\bigg|_{\omega\rightarrow 0}\frac{2k_BT}{\hbar},
\end{split}
\end{equation}
where $e$ is the elementary charge, $\Psi_{\uparrow}$ and $\Psi_{\downarrow}$ are the electron spin-up and spin-down molecular wavefunctions solved by the atomistic tight-binding method respectively, $\omega$ is the noise frequency and $S_E(\omega)$ is the noise field spectrum. For $S_E(\omega)$, we study $1/f^{\alpha}$ noise, Johnson noise and evanescent wave Johnson noise (EWJN) \cite{PHuang}, assuming the noise source is 65 nm (the distance between G1 and the two dot center) away from the qubit system to be consistent with Ref. \cite{Tom_readout}. The expressions of the noise field spectra and the parameter estimations based on experiments \cite{Tom_readout,Bent_prl,Buch_prb} are also included in the Supplementary.

\begin{figure}[h]
\center\epsfxsize=3.4in\epsfbox{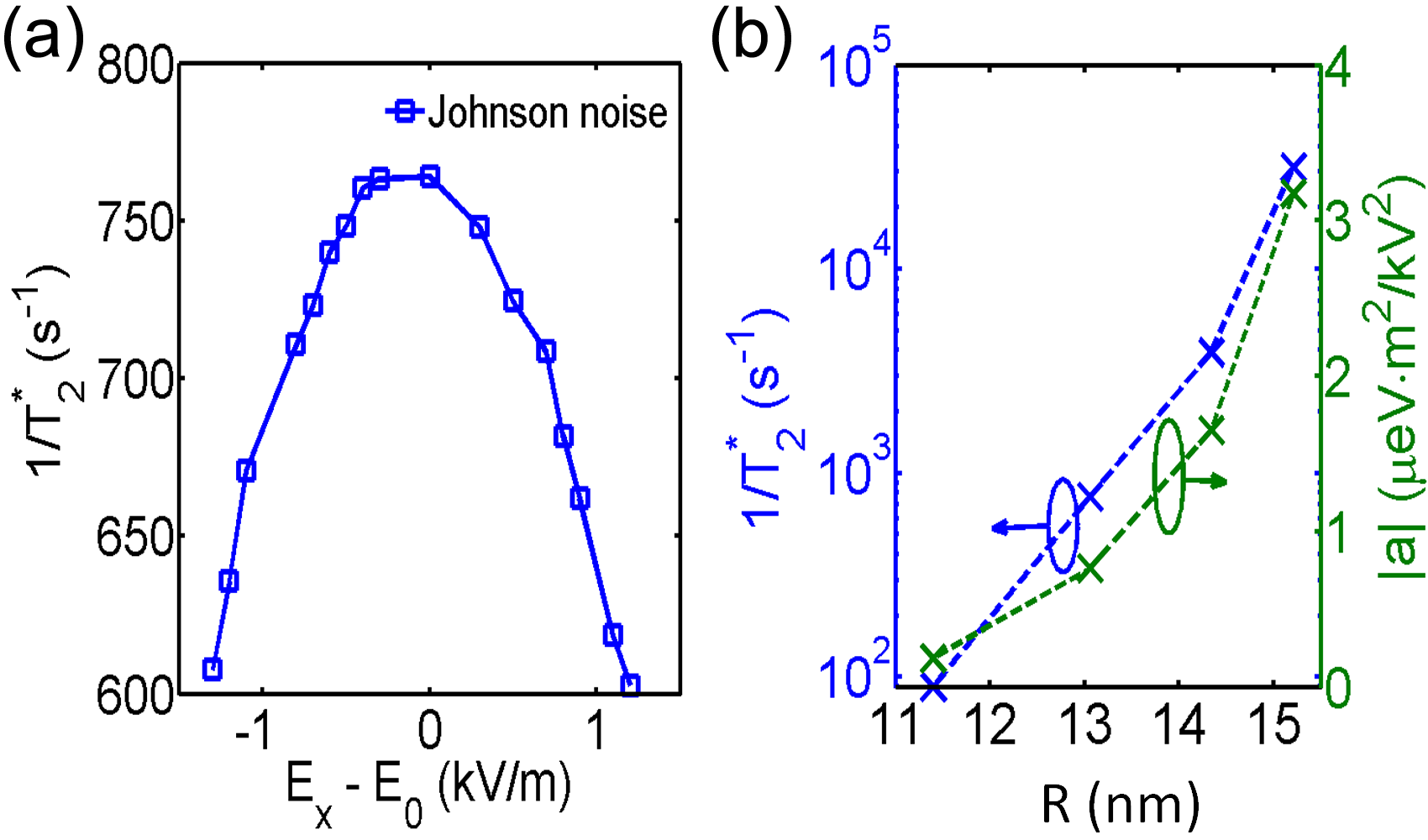}
\caption{ \textbf{Impact of electric field and inter-dot separation on decoherence time.} (a) Qubit decoherence rate $1/T_{2}^{*}$ as a function of dc electric field in the x direction due to Johnson noise from a single noise source, for R $\approx$ 13 nm and B = 0.5 T. Here, $E_0$ = 5682.2 kV/m. (b) The maximum $1/T_{2}^{*}$ due to Johnson noise (left y-axis) and energy curve curvatures ($|a|$ on the right y-axis) as a function of 2P-1P inter-dot separation $R$.}
\vspace{0cm}
\label{fig_noise}
\end{figure}

Fig. 4(a) demonstrates the decoherence rate $1/T_{2}^{*}$ as a function of the applied dc electric field ($E_x$) due to Johnson noise from a single noise source for the case $R \approx$ 13 nm, assuming B = 0.5 T. Using the estimated parameters in the Supplementary, we find that $T_{2}^{*}$ is limited by Johnson noise. Based on our calculations, the effect of EWJN is at least 2 orders of magnitude lower than Johnson noise, and $1/f^{\alpha}$ noise is negligible (see Supplementary), thus they are not shown here. As shown, the decoherence rate $1/T_{2}^{*}$ reaches a maximum at $E_x=E_0$, where the bonding and anti-bonding states are formed and the system is most sensitive to charge noise. 
The left y-axis of Fig. 4(b) shows the maximum decoherence rate $1/T_{2}^{*}$ due to Johnson noise as a function of inter-dot separation $R$. As shown, $T_{2}^{*}$ can be improved by shrinking the inter-dot separation. This can be explained by the curvature of the qubit energy curve (e.g. the solid green or blue curve in Fig. 3(a))), which serves as a metric of how the qubit is prone to charge noise near $E_x = E_0$. On the right y-axis of Fig. 4(b), we plot this curvature ($a$ in its absolute value) by fitting the energy curves with a quadratic function of $E_x$ for different $R$. The curvature term $|a|$ increases with $R$ because the tunnel coupling $t_c$ decreases with $R$ (Fig. 3(c)), causing more abrupt charge transition. As can be seen, $1/T_2^{*}$ agrees with the trend of the curvature term $|a|$. Consequently, larger tunnel coupling/smaller 2P-1P separation is preferred to enhance the qubit coherence time.

So far, we have investigated the decoherence time due to a single charge noise source. In a real device, there could be multiple charge noise sources (other in-plane gates, top gates, etc.), resulting in $T_2^{*}$ being degraded by 1-2 orders of magnitude eventually. Even then, using a 2P-1P spin qubit with electrical control is likely to yield devices comparable to single electron spin qubit based on single donor ($T_2^*$ = 268 $\mu$s \cite{Muhonen}) and single quantum dot qubit ($T_2^*$ = 120 $\mu s$ \cite{Menno1}) based on magnetic control, and outperform the 1P-1P charge qubit ($T_2^*$ = 0.72 $\mu$s \cite{Hollenberg}) in terms of qubit coherence time. 

\section{Summary}
In summary, we propose a novel approach for all-electrical control of donor-based spin qubits in silicon using full-band atomistic tight-binding modeling and time-dependent simulations based on effective spin Hamiltonian. In this design, ancillary dots are introduced to form an asymmetric 2P-1P system to create a hyperfine coupling difference between 2P and 1P, utilized to realize electron spin control with an ac electric field. We perform a quantitative analysis to optimize this design in terms of X-gate fidelity and decoherence time through external static B-field and tunnel coupling determined by inter-dot separation. We show that a long qubit coherence time can be potentially achieved. This work can serve as an alternative design to those that exploit the hyperfine difference between the donor and the interface states \cite{Tosi}, where the qubit coherence could be affected by the proximity of the oxide interface \cite{Schenkel,Paik}. To further reduce possible sources of deleterious noise in the proposed design, we would further pursue all-in-plane electrostatic and qubit control without the top surface gates in the future.  

\section*{Acknowledgements}
This research was conducted by the Australian Research Council Centre of Excellence for Quantum Computation and Communication Technology (project No. CE110001027), the US National Security Agency and the US Army Research Office under contract No. W911NF-08-1-0527. Computational resources on nanoHUB.org, funded by the NSF grant EEC-0228390, were used. M.Y.S. acknowledges a Laureate Fellowship. 

\section*{Supplementary}

{\textit{Spectrum functions of charge noise fields.}} Following Ref. \cite{PHuang}, we investigate three types of charge noise. The $1/f^{\alpha}$ noise field spectrum is expressed as:
\begin{equation}
S_E(\omega) = \frac{N}{\omega^{\alpha}},
\end{equation} 
where $N$ is the noise field strength in $(V/m)^2$. In this work, we estimate $N$ based on Ref. {\cite{Kuhlmann}}, where the root-mean-square electric field noise is $F_{r.m.s}$ = 46 V/m. Then $N = (\sqrt{2}F_{r.m.s})^2$ = 4232 $(V/m)^2$. We assume the bandwidth starts at 0.1 Hz also in line with Ref. {\cite{Kuhlmann}}, and $\alpha$ = 1 in the calculations.

The Johnson noise field spectrum is:
\begin{equation}
S_E(\omega) = \frac{2\xi\omega\hbar^2}{1+(\omega/\omega_R)^2}/(el_0)^2,
\end{equation}
where $\xi = R/R_k$, $R_k$ is the fundamental quantum resistance $h/e^2$, $R$ is the circuit resistance, and $\omega_R = 1/RC$ is the cutoff frequency. R is estimated based on Ref. \cite{Bent_prl}. In this work, we assume the gate length ($l_g$) is 100 nm and the gate width is 6 nm ($w$) which leads to 18 conducting modes ($M$, number of modes) \cite{Bent_prl}. As estimated in Ref. \cite{Bent_prl}, the mean-free-path ($\lambda$) of such a wire is $\sim$6 nm. Hence, R in this work is calculated by $1/R = e^2/h\cdot M\cdot \lambda/(\lambda+l_g)$. $l_0$ is the distance between the qubit and the noise source, and we assume $l_0$ = 65 nm based on experimental devices \cite{Tom_readout}. $C$ is estimated based on Ref. \cite{Buch_prb}, where the donor-gate capacitance is 0.6 aF for a separation $\sim$35 nm. Therefore, $C$ in this work is evaluated as $35nm/65nm\cdot 0.36 aF = 0.17 aF$.

The evanescent wave Johnson noise (EWJN) field spectrum can be expressed as:
\begin{equation}
S_E(\omega) = \frac{\hbar\omega}{8z^3\sigma},
\end{equation}
where $\sigma$ is the conductivity of the gates. We extract $\sigma$ from Ref. \cite{Bent_prl} where the wire length ($l_w$) is 47 nm. We assume the thickness of the wire ($t_w$) to be 2 nm considering donor diffusion and segregation, then the conductivity can be expressed as $4.8\frac{e^2}{h}\frac{l_w+\lambda}{w\cdot t_w}$.

\begin{figure}[htbp]
\renewcommand\thefigure{S1}  
\centering\epsfxsize=3.4in\epsfbox{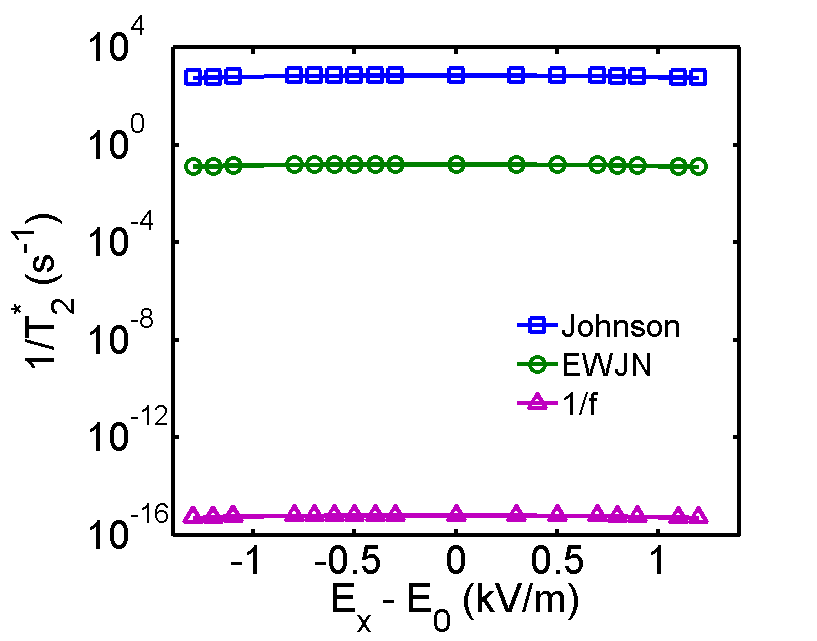}
\caption{ Comparison of the effects of Johnson noise, evanesant wave Johnson noise and $1/f$ noise on the decoherence rate $1/T_2^{*}$. The 2P-1P separation $R\approx$ 13 nm and the applied static B-field is 0.5 T.}
\label{fis}
\end{figure}

In Fig. {\ref{fis}}, we compare the effects of the three types of charge noise stated above on the decoherence rate {$1/T_2^{*}$}. As can be seen and stated in the main text, {$T_{2}^{*}$} is limited by Johnson noise. EWJN is at least 2 orders of magnitude lower than Johnson noise. {$1/f^{\alpha}$} noise is negligible, which is consistent with Ref. {\cite{PHuang}}.

Electronic address: wang1613@purdue.edu; rrahman@purdue.edu


\begin{thebibliography}{100}
\bibitem{charge_config_note} The charge configuration of the system is denoted as (M,N), where M indicates the number of electrons on the left dot, and N the right dot.
\end{thebibliography}
\end{document}